\documentclass{article}

\usepackage[preprint]{neurips_2026}

\usepackage[utf8]{inputenc}
\usepackage[T1]{fontenc}
\usepackage{hyperref}
\usepackage{url}
\usepackage{booktabs}
\usepackage{amsfonts}
\usepackage{amsmath}
\usepackage{amssymb}
\usepackage{nicefrac}
\usepackage{microtype}
\usepackage{xcolor}
\usepackage{colortbl}
\definecolor{fullrowblue}{RGB}{220,235,250}
\usepackage{graphicx}
\usepackage{subcaption}
\usepackage{multirow}
\usepackage{pifont}
\usepackage{algorithm}
\usepackage{algorithmic}
\usepackage{enumitem}
\usepackage{wrapfig}
\usepackage{placeins}
\newcommand{\mesh}{\mathcal{M}}
\newcommand{\context}{\mathcal{A}\!\setminus\!\mathcal{B}}
\newcommand{\target}{\mathcal{B}}
\newcommand{\generated}{\mathcal{B}'}
\newcommand{\pointcloud}{\mathbf{P}}

\usepackage{soul}            

\newcommand{\cmark}{\ding{51}}
\newcommand{\xmark}{\ding{55}}

\usepackage{fancyhdr}
\AtBeginDocument{%
  \addtolength{\headheight}{6pt}%
  \addtolength{\headsep}{-6pt}%
}
\fancypagestyle{hunyuanheader}{
    \fancyhf{}
    \lhead{%
        \raisebox{-0.2\height}{\includegraphics[height=12pt]{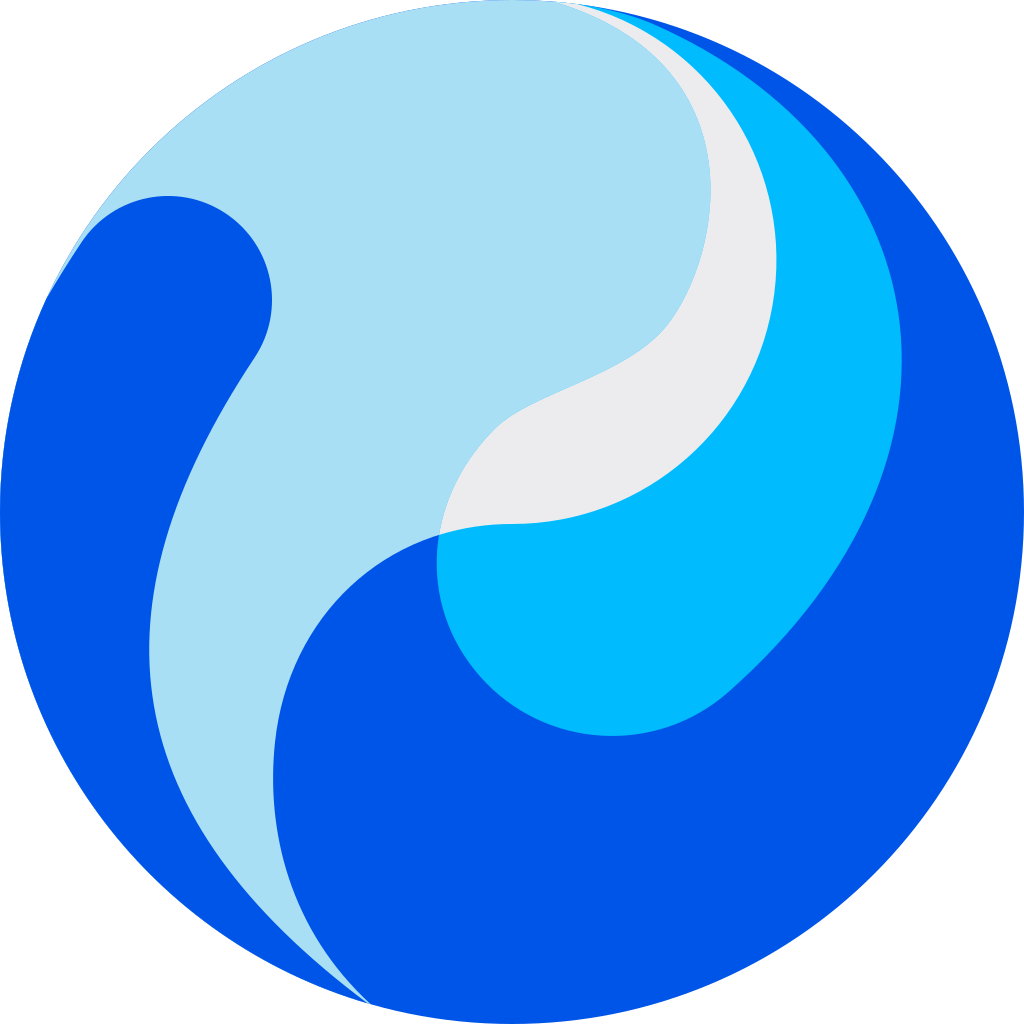}}%
        \hspace{3pt}%
        \textsf{Tencent Hunyuan}%
    }
    \rhead{\textsf{Preprint}}
    
}

\title{MeshFIM: Local Low-Poly Mesh Editing via Fill-in-the-Middle Autoregressive Generation}

\author{\mdseries
  \textbf{Dingdong Yang}\textsuperscript{1,2}\quad
  \textbf{Jian Liu}\textsuperscript{1,3}\quad
  \textbf{Biwen Lei}\textsuperscript{1}\quad
  \textbf{Haohan Wang}\textsuperscript{1}\quad
  \textbf{Zhuo Chen}\textsuperscript{1}\\
  \textbf{Song Guo}\textsuperscript{3}\quad
  \textbf{Ali Mahdavi-Amiri}\textsuperscript{2}\quad
  \textbf{Hao Zhang}\textsuperscript{2,$\dagger$}\quad
  \textbf{Chunchao Guo}\textsuperscript{1}\\[6pt]
  \textsuperscript{1}Tencent Hunyuan\quad
  \textsuperscript{2}Simon Fraser University\quad
  \textsuperscript{3}Hong Kong University of Science and Technology\\[3pt]
  {\small \textsuperscript{$\dagger$}Corresponding author.}
}

\begin{document}

\maketitle
\thispagestyle{hunyuanheader}

\vspace{-0.5em}
\begin{center}
    \includegraphics[width=\textwidth]{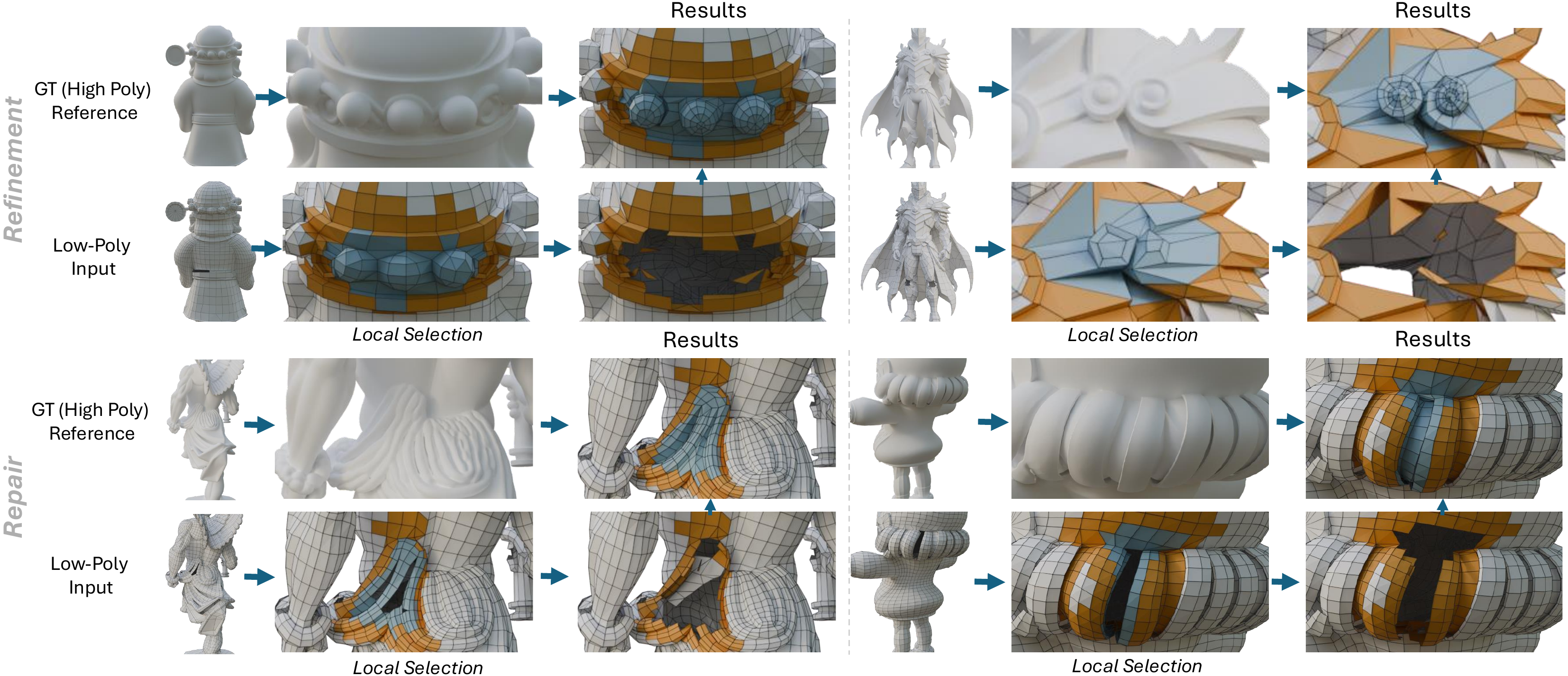}
    \captionof{figure}{
    \textbf{MeshFIM} enables accurate local low-poly mesh editing.
    Given an input low-poly mesh, a target local region
    (\textcolor[RGB]{126,143,158}{blue}) is selected via flexible interactive tools (e.g.~sec.\ref{ssec:brush}) 
    or automatic defect detection algorithms (e.g.~sec.\ref{ssec:hole-repair}).
    Conditioned on ground truth reference high poly and context mesh (\textcolor[RGB]{174,126,75}{orange}), 
    MeshFIM regenerates the target region for refinement (top) or repair (bottom) while
    preserving seamless boundary connectivity and smooth edge flow 
    with surrounding existing mesh. 
  }
    \label{fig:teaser}
\end{center}
\vspace{-0.5em}

\begin{abstract}
Autoregressive (AR) models can generate high-quality low-poly meshes
from point clouds, but they still operate in an all-or-nothing
manner: when a local region is unsatisfactory, the entire mesh must
be regenerated, wasting computation and destroying satisfactory
mesh structure elsewhere.
We introduce \textbf{MeshFIM}, a Fill-in-the-Middle (FIM) framework
that regenerates a target region of a low-poly mesh conditioned on the
surrounding context.
MeshFIM addresses three mesh-specific challenges:  enforcing exact attachment along
 the exposed boundary, preserving topological order in the context, and suppressing overflow beyond the intended
 region.
It does so with five complementary design choices:  boundary vertex markers, context positional
embeddings, expanded context width, context
augmentation, and a low-poly geometry encoder whose gated subtraction
mechanism focuses generation on the missing region by leveraging the
difference between the reference surface and the existing mesh. Detailed ablation studies are presented to show the effectiveness of every introduced component.
Based on MeshFIM, we demonstrate two applications: interactive brush-based editing and
automatic defect repair on low-poly mesh (see~\autoref{fig:teaser}). 
Last but not least, experiments show that MeshFIM outperforms a range of baselines in mesh refinement, 
mesh repair and whole mesh generation plus stitch-back scheme. 
\end{abstract}

\section{Introduction}
\label{sec:intro}

Geometric models in practice are rarely artifact-free or remain a finished product. Meshes acquired via scanning 
carry artifacts from occlusion and sensor noise, while those generated by AI 
can exhibit malformed local topology and density irregularities. Even well-formed meshes must evolve as their usage scenarios change: a game asset needs denser topology near a newly rigged joint, a CAD model requires topological cleanup before simulation, 
while real-time applications often demand 3D objects to adapt to different levels of detail. 
In many such cases, the required mesh editing 
is \emph{local}, with a small region to be repaired, refined, or rewired while the rest of the mesh remains intact.

Classical local mesh editing 
via hole filling, 
Poisson reconstruction, 
or subdivision focuses 
on geometric criteria such as surface smoothness and shape approximation; they are often oblivious to the quality of mesh topology or connectivity, especially when accounting for the topological context of the surrounding mesh or the semantic structure of the edited object. This limitation is especially acute in the \emph{low-poly} setting, where mesh topology is not merely a surface discretization but encodes design intent. This motivates learning-based approaches, trained on large collections of artist-made meshes, to enable editing that is both geometrically valid and topologically coherent.

Recently proposed autoregressive (AR) mesh generators can produce artist-grade low-poly meshes directly from geometric inputs such as point clouds. Models such as MeshGPT~\cite{siddiqui2024meshgpt}, MeshAnything~\cite{chen2024meshanything,chen2024meshanythingv2}, QuadGPT~\cite{liu2026quadgpt}, 
among others~\cite{hao2024meshtron, chen2024meshxl, zhao2025deepmesh, tang2025edgerunner,lei2025ARMesh,weng2025BPT,lionar2025treemeshgpt}, generate meshes token-by-token, attaining high-quality mesh topology.
However, to our knowledge, existing AR generators all operate in an \emph{all-or-nothing} manner. That is, when only a local region needs adjustment, an entire mesh is regenerated from scratch. This is wasteful, at odds with practical modeling workflows, and can potentially destroy satisfactory topology elsewhere.


We introduce MeshFIM, a local mesh editor built on the \emph{Fill-in-the-Middle} (FIM) paradigm from code 
completion~\citep{fried2022incoder,guo2024deepseek}. Given a target region designated for editing, MeshFIM \emph{autoregressively} generates replacement faces conditioned on, and seamlessly attached to, the surrounding mesh as the \emph{topological context}, analogous to how code FIM predicts a masked token span from its prefix to its suffix. However, meshes are two-dimensional complexes embedded in three dimensions rather than one-dimensional token streams. Hence, directly applying the FIM paradigm to meshes introduces non-trivial geometric and topological constraints that require careful treatment.


Specifically, a challenge unique to meshes is that when the surrounding context faces are serialized into a flat token sequence, as required by any AR generator, a spatially contiguous neighborhood loses its local ordering, making it difficult for the model to preserve topological patterns such as edge flow and density transitions. Also critically, the generated patch must reuse the exposed boundary vertices exactly: even sub-quantization coordinate mismatches create visible cracks or T-junctions at the seam. Finally, at inference time, the surrounding context may already contain artifacts from prior generation steps or imperfect user selections, whereas training contexts are extracted from clean ground-truth meshes, leading to a train-inference gap that degrades robustness in practice.


MeshFIM addresses all these challenges within a general AR mesh generation framework:
\begin{itemize}
\item To handle the serialization ordering problem, we introduce \emph{context positional embeddings} that restore local topological structure within the flattened context sequence. 
\item To enforce exact boundary attachment, learned boundary vertex markers explicitly signal to the decoder which context vertices must be reused as stitching anchors. 
\item To prevent overflow (i.e., generated faces spilling beyond the intended region) and improve generation quality, we introduce two complementary techniques:
(1) an expanded context window exposes sufficient surrounding geometry to keep generation local; 
(2) a low-poly geometry encoder with gated subtraction continuously signals where the existing mesh already covers the reference surface.
\item Finally, to close the train-inference gap, we introduce context augmentation which, during training, improves robustness to noisy and incomplete contexts at inference time.
\end{itemize}

Our method is comprehensively evaluated against various baselines~\autoref{ssec:main-results} and thorough ablation studies~\autoref{ssec:ablation} are presented 
to validate the effectiveness of our algorithm components.

\section{Related Work}
\label{sec:related}

\paragraph{Autoregressive Mesh Generation.}
PolyGen~\cite{nash2020polygen} first demonstrated vertex-then-face autoregressive
mesh generation. MeshGPT~\cite{siddiqui2024meshgpt} introduced a VQ-VAE tokenizer
followed by a GPT-style decoder. MeshAnything~\cite{chen2024meshanything} and
MeshAnythingV2~\cite{chen2024meshanythingv2} scaled the approach to artist-quality meshes.
Many follow-up methods further improved tokenization and training strategies~\cite{chen2024meshxl,tang2025edgerunner,weng2025BPT,lionar2025treemeshgpt,lei2025ARMesh,zhao2025deepmesh}.
More recently, QuadGPT~\cite{liu2026quadgpt} and
Meshtron~\cite{hao2024meshtron} extended the paradigm to
quad-dominant meshes by directly generating mixed quad-triangle faces
in a canonical lexicographic order. All these methods generate
\emph{complete} meshes, some in the low-poly setting, but none support local low-poly mesh editing.

\vspace{-5pt}

\paragraph{Local Subdivision.}
Local subdivision~\citep{catmull1998recursively,zorin1996interpolating,loop1987smooth,kobbelt20003,liu2020neural,zhu2025neural} has been a standard approach to
increasing mesh density in a target region. However, these methods can only \emph{add} (split) faces, and are not designed to rewire mesh topology. In contrast, MeshFIM regenerates topology from scratch in a local region.

\vspace{-5pt}

\paragraph{Mesh Completion and Repair.}
Classical hole-filling algorithms~\citep{liepa2003filling, tekumalla2004hole, ju2004robust, kazhdan2006poisson, zhao2007robust, kazhdan2013screened, centin2015poisson} operate on
triangle soups and do not preserve the original face structure.
Optimization-based mesh repair~\citep{hattori2024learning} lacks generalization capabilities and is unable to repair complicated missing regions. Conditioned on a reference dense mesh and a surrounding context mesh, our approach regenerates topology in a local region using a learned AR mesh model; it is general and perfectly aligned with surrounding mesh faces.

\vspace{-5pt}

\paragraph{Fill-in-the-Middle for Sequences.}
FIM training was introduced for code language models~\citep{bavarian2022efficient,lewis2020bart,roziere2023code,guo2024deepseek},
enabling infilling of contiguous spans. InCoder~\cite{fried2022incoder} and
subsequent work showed that FIM can be added to pretrained models with
minimal degradation of left-to-right performance. However, since mesh resides in 3D space, 
its serialization token sequence depends on specific mesh topology, see~\autoref{fig:token-sequence}, 
namely where to put the missing tokens is unknown before eventual generation.


\section{Method}
\label{sec:method}

\subsection{Preliminaries: Autoregressive Mesh Generation}
\label{ssec:preliminaries}

\begin{figure}[t]
    \centering
    \includegraphics[width=\linewidth]{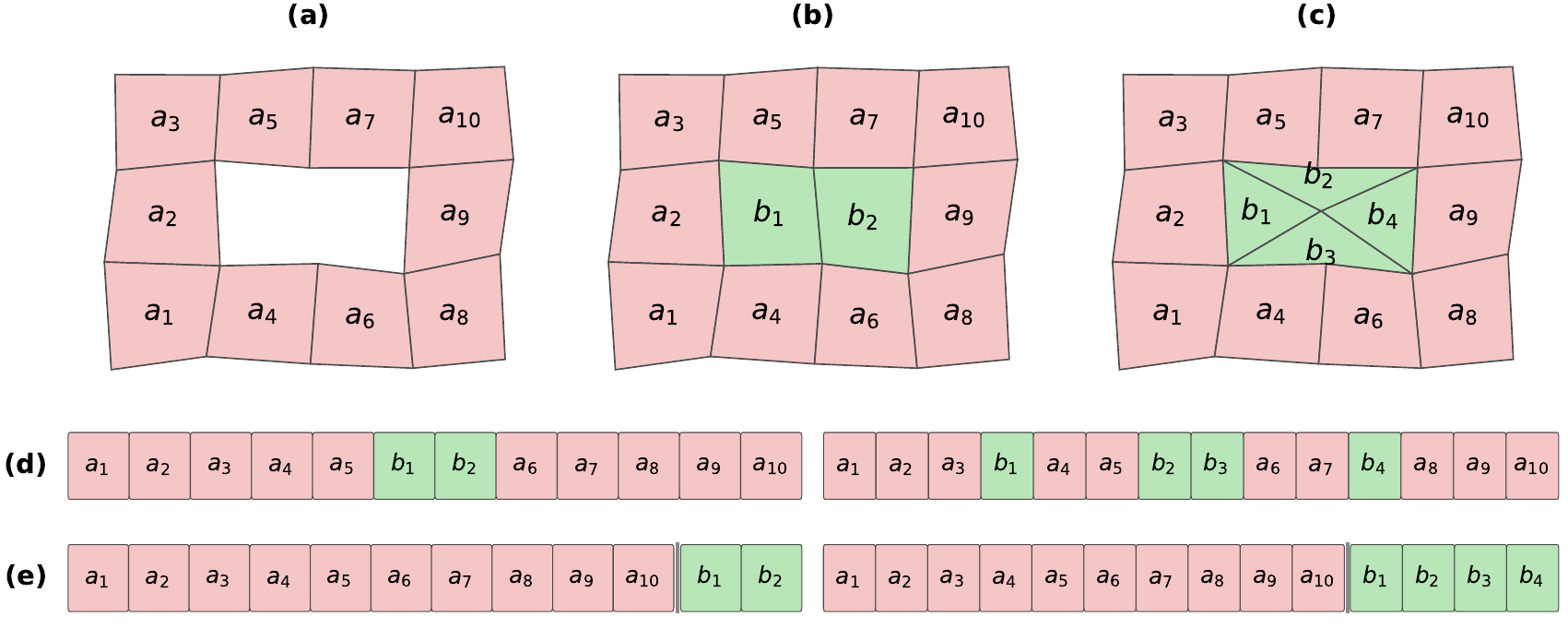}
    \caption{\textbf{Naive global serialization conflicts with local
    editing} (faces sorted in YX lexicographic order).
    (a)~A fixed context $\context$.
    (b,\,c)~Two valid tilings of the same hole with different face
    counts and shapes.
    (d)~Under global serialization, the positions and number
    of target tokens
    ($b_i$, \textcolor[HTML]{5BAF5B}{green}) in the joint sequence are
    determined by the target geometry itself, which is unknown before
    generation: different tilings produce different sequences.
    (e)~MeshFIM serializes $\context$
    (\textcolor[HTML]{D98C8C}{pink}) and $\target$
    (\textcolor[HTML]{5BAF5B}{green}) independently, so that the
    sequence structure no longer depends on the unknown target geometry.}
    \label{fig:token-sequence}
\end{figure}

MeshFIM builds on recent autoregressive mesh
generators introduced by~\cite{liu2026quadgpt,hao2024meshtron} that formulate mesh
generation as next-token prediction. A mesh $\mesh$ with $F$ faces is
serialized into a token sequence so that each face $f_j$
($j=1,\dots,F$) is represented by $n$ vertices ($n\!=\!4$ for quads,
$n\!=\!3$ for triangles with padding to a uniform length), and each
vertex is encoded as three quantized coordinate tokens, yielding a
fixed-length block $\mathbf{b}_j \in \mathbb{Z}^{12}$ per face.
Faces are sorted in a canonical bottom-to-top lexicographic order
(by the YXZ coordinates of their lowest vertex), producing a
deterministic sequence
$\mathbf{x} \;=\; \mathbf{b}_1 \oplus \mathbf{b}_2 \oplus
    \cdots \oplus \mathbf{b}_F\,$,
where $\oplus$ denotes concatenation. An Hourglass
Transformer~\cite{nawrot2022hierarchical} then models
\begin{equation}
    p(\mesh \mid \pointcloud) \;=\;
    \prod_{i=1}^{|\mathbf{x}|}
    p(x_i \mid x_{<i},\; \pointcloud),
    \label{eq:ar-objective}
\end{equation}
autoregressively predicting each coordinate token $x_i$ conditioned
on preceding tokens and a reference geometry (point-cloud $\pointcloud$).
Following prior work, $\pointcloud$ is injected by a
Perceiver-IO-style encoder with cross-attention queries and
self-attention blocks~\cite{jaegle2021perceiver,zhang20233dshape2vecset,zhao2023michelangelo}.
This formulation is effective for whole-mesh generation but not directly suited for local editing. 
A connected target region on the surface
generally becomes scattered after serialization
(\autoref{fig:token-sequence}), and standard
left-to-right decoding does not enforce exact reuse of boundary
vertices when a regenerated patch is stitched back into the remaining
mesh. These issues motivate MeshFIM.

Given this autoregressive representation, we cast local mesh editing
as conditional infilling. The rest of this section describes the
formulation~(\autoref{ssec:formulation}) and the core design
choices~(\autoref{ssec:arch}).

\subsection{Problem Formulation}
\label{ssec:formulation}

\begin{figure}[t]
    \centering
    \includegraphics[width=\linewidth]{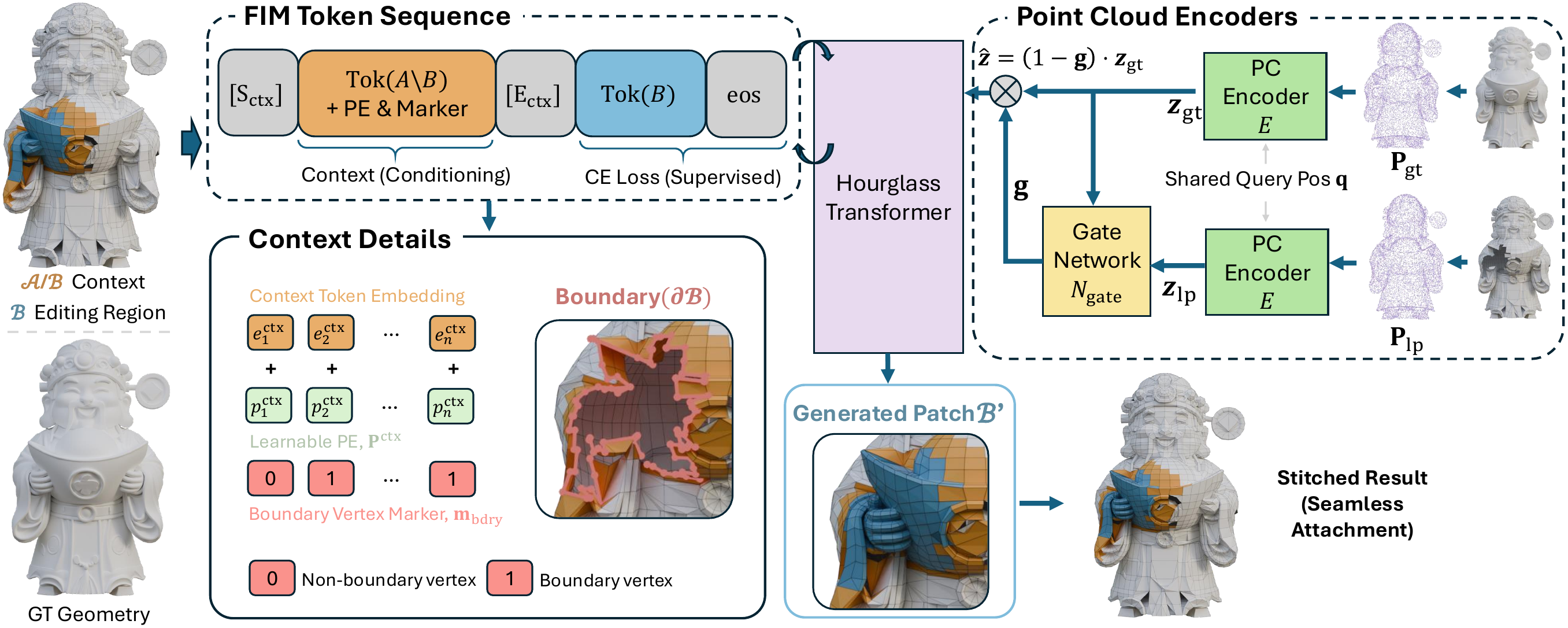}
    \caption{\textbf{MeshFIM pipeline overview.}
    Given an input mesh with a \textcolor[RGB]{126,143,158}{target region} $\target$ and
    surrounding \textcolor[RGB]{174,126,75}{context} $\context$, a shared point cloud encoder $E$
    with a common set of query positions $\mathbf{q}$ encodes two
    point clouds---the reference one $\mathbf{P}_{\text{gt}}$ and
    the existing low-poly mesh $\mathbf{P}_{\text{lp}}$ (with
    $\target$ removed)---into position-aligned latents
    $\mathbf{z}_{\text{gt}}$ and $\mathbf{z}_{\text{lp}}$.
    The mesh is serialized into a FIM token sequence:
    context tokens (with positional embeddings and boundary markers)
    followed by target tokens.
    An Hourglass Transformer autoregressively generates the new
    patch $\generated$, conditioned on the fused latents $\hat{\mathbf{z}}$ via
    cross-attention.
    The output is stitched back seamlessly along the exposed boundary.
    }
    \label{fig:pipeline}
\end{figure}

Let $\mesh$ denote an input low-poly mesh, and let $\target \subset
\mesh$ denote a target region of interest (blue region in
\autoref{fig:pipeline}). We remove the target faces and define the
\emph{context} $\context$ (orange region in \autoref{fig:pipeline}) as
the set of faces within $w$ breadth-first rings of $\target$. The
objective is to learn a conditional distribution
\begin{equation}
    p\!\left(\generated \mid \context,\; \pointcloud \right),
    \label{eq:fim-objective}
\end{equation}
where $\generated$ denotes the generated faces that complete the
region while stitching seamlessly to $\context$ and faithfully
following the reference geometry $\pointcloud$. Concretely, MeshFIM serializes the context first and then decodes the
replacement patch autoregressively:
\begin{equation}
    [\mathrm{S}_{\mathrm{ctx}}] \oplus \operatorname{Tok}(\context)
    \oplus [\mathrm{E}_{\mathrm{ctx}}] \oplus
    \operatorname{Tok}(\generated) \oplus [\mathrm{eos}],
    \label{eq:fim-sequence}
\end{equation}
where $[\mathrm{S}_{\mathrm{ctx}}]$ and $[\mathrm{E}_{\mathrm{ctx}}]$
are learned start and end sentinels, $\operatorname{Tok}(\cdot)$
denotes face tokenization, and $[\mathrm{eos}]$ ends the
sequence. During training, the cross-entropy loss is applied only to
$\operatorname{Tok}(\generated)$ and $[\mathrm{eos}]$, so the context
serves as a conditioning rather than a prediction
target. Note that during training, $\generated = \target$.

\subsection{Core Design Choices}
\label{ssec:arch}

MeshFIM adopts five complementary design choices. 
We first make boundary attachment constraints explicit by \textbf{boundary vertex markers}, 
then \textbf{expand context width} 
to reduce generation overflow, 
add \textbf{context positional embeddings} 
to recover context topology, 
and \textbf{context augmentation} 
during training to close the train-inference gap arising from noisy inference-time inputs,
and finally introduce a \textbf{low-poly geometry encoder} 
to focus on generation over missing regions.

%
\paragraph{Boundary vertex markers.}
At the interface between $\context$ and $\generated$, even small
coordinate mismatches create cracks or T-junctions. We therefore mark
all context tokens that correspond to vertices on the exposed
boundary. Let $V_{\partial}$ denote this set. Their
embeddings are modified as
\begin{equation}
    \tilde{\mathbf{e}}_i^{\mathrm{ctx}} \;=\;
    \mathbf{e}_i^{\mathrm{ctx}} + \mathbf{m}_{\mathrm{bdry}},
    \quad \text{if } t_i \in V_{\partial},
    \label{eq:shared-vtx}
\end{equation}
where $\mathbf{m}_{\mathrm{bdry}} \in \mathbb{R}^d$ is a learned
marker. It is an explicit cue informing the decoder which context vertices
must be reused as stitching anchors, and shown to be essential for reliable 
boundary matching; see our ablation studies in
Table~\ref{tab:ablation}.

%
\paragraph{Expanded context width.}
A narrow context provides only minimal information about local
curvature and edge flow, which often causes \emph{overflow
generation}: newly generated faces spilling beyond the intended region.
Using a wider context exposes more of the surrounding surface to ease
geometry inference. In practice, three
breadth-first rings provide a good trade-off between geometric
coverage and prompt length. As shown in Figure~\ref{fig:overflow} and
Table~\ref{tab:ablation}, a wider context sharply reduces overflow.

\vspace{-5pt}

%
\paragraph{Context positional embeddings.}
Beyond seamless stitching, a plausible edit should inherit local face
density and edge flow from its neighborhood. If the context is given
only as a flat token sequence, the model can treat it as an unordered
bag of nearby vertices and lose the structure needed to preserve those
patterns. We therefore add a learned positional encoding over the
serialized context sequence:
\begin{equation}
    \mathbf{e}_i^{\mathrm{ctx}} \;=\;
    \operatorname{Emb}(t_i^{\mathrm{ctx}}) + \mathbf{p}_i^{\mathrm{ctx}},
    \quad i = 1, \ldots, |\operatorname{Tok}(\context)|,
    \label{eq:ctx-pos-emb}
\end{equation}
where $\mathbf{p}_i^{\mathrm{ctx}}$ is a learnable positional code.
This restores order within the prompt and helps the model preserve the
surrounding topological pattern. Qualitative evidence is shown in
\autoref{fig:ctx-pos-emb}, and the corresponding ablation study can be found in
Table~\ref{tab:ablation}.

\vspace{-5pt}

\paragraph{Low-poly geometry encoder.}
The point-cloud conditioner encodes the reference surface
(what the mesh \emph{should} look like) but has no
knowledge of what the current low-poly mesh \emph{already} covers.
This information reaches the decoder only indirectly through the FIM context tokens. We therefore reuse the same Perceiver-IO
encoder $E$ to encode two point clouds, $\pointcloud_{\text{gt}}$
sampled from the reference surface and $\pointcloud_{\text{lp}}$
sampled from the existing low-poly mesh with $\target$ removed, under
a \emph{shared} set of cross-attention query positions
$\mathbf{q}=\{\mathbf{q}_i\}_{i=1}^{M}$ drawn in Euclidean space:
\begin{equation}
    \mathbf{z}_{\text{gt}} = E(\pointcloud_{\text{gt}}, \mathbf{q}),
    \qquad
    \mathbf{z}_{\text{lp}} = E(\pointcloud_{\text{lp}}, \mathbf{q}).
    \label{eq:lp-encode}
\end{equation}
Sharing $\mathbf{q}$ aligns the two latent sequences position by
position, so a per-slot gate is well-defined.

A lightweight gate network $N_{\mathrm{gate}}$ produces a vector
$\mathbf{g} \in (0, 1)^M$ from the two aligned latents, and the fused
conditioning is an attenuation of $\mathbf{z}_{\text{gt}}$:
\begin{equation}
    \mathbf{g} = 
    \sigma\!\bigl(N_{\mathrm{gate}}([\mathbf{z}_{\text{gt}};\,
      \mathbf{z}_{\text{lp}}])\bigr),
    \qquad
    \hat{\mathbf{z}} = (1 - \mathbf{g})\,\mathbf{z}_{\text{gt}},
    \label{eq:lp-gate}
\end{equation}
where $[\,\cdot\,;\,\cdot\,]$ denotes channel-wise concatenation of the
two position-aligned latents at the shared query $\mathbf{q}$,
$\sigma$ is the sigmoid function, and $N_{\mathrm{gate}}$ is
initialized so that training starts from
$\hat{\mathbf{z}}\approx\mathbf{z}_{\text{gt}}$ (architecture and
initialization in Appendix~\ref{app:gate}).
Intuitively, where the existing mesh already covers the reference
surface well, $g_i$ is large and the reference conditioning at
$\mathbf{q}_i$ is suppressed; where the mesh is missing (the hole),
$g_i$ stays small and the decoder sees the full reference latent,
directing generation toward the missing region.
This provides a continuous, spatially grounded signal that complements
the discrete token context for overflow suppression and region focusing
(Table~\ref{tab:ablation}, Figure~\ref{fig:gate-vis}).

\vspace{-5pt}

\paragraph{Context augmentation.}
During training, contexts are extracted from ground-truth meshes, which
are cleaner than those encountered at inference that may already contain quantization noise or artifacts from earlier
generation steps. To narrow this gap, we perturb context vertex coordinates by random
offsets drawn from $\mathcal{U}(-\delta, \delta)$. Shared boundary vertices $\partial\target$ are also perturbed. This exposes the
model to imperfect boundaries during training to improve robustness, as confirmed by Table~\ref{tab:ablation}.

\section{Experiments}
\label{sec:experiments}
\vspace{-2pt}

\subsection{Setup}
\label{ssec:setup}

\paragraph{Implementation Details.}
We use the AR mesh generator described in
\autoref{ssec:preliminaries}: an Hourglass
Transformer~\cite{nawrot2022hierarchical} ($d\!=\!1536$, depth 24,
16 heads) conditioned on point clouds encoded by an
encoder~\cite{jaegle2021perceiver,zhao2023michelangelo}. 
We train the model on 64 computing units, each with about 90 GB memory.
Context augmentation uses $\mathcal{U}(-0.002,0.002)$ noise.
Target regions $\target$ are sampled on the face
adjacency graph with up to 1200 faces in a BFS-driven strategy.
To expose the model to irregular editing targets that better resemble
brush selections as well as automatically detected defects, 30\% of training
samples use a percolation-based region growth strategy: starting from
a random seed face, the region expands stochastically along the
frontier with acceptance probability
$p\!\sim\!\mathcal{U}(0.55,0.85)$ until the face budget is reached
(see~\autoref{app:percolation} for the full algorithm).

\vspace{-5pt}

\paragraph{Baselines.} 
To our knowledge, MeshFIM makes the first attempt at accurate local low-poly mesh editing/generation, hence there is no clearly suitable method to compare with.
In making a best effort and considering the potential applications, we consider the following as comparable baselines:
\begin{enumerate}
\item Mesh Refinement: Modified Butterfly~\cite{zorin1996interpolating},
Neural Subdivision~\cite{liu2020neural}, and Neural Mesh Refinement~\cite{zhu2025neural}.
\item Mesh Repair: Liepa Hole Filling~\cite{liepa2003filling} and
an optimization-based mesh repair method (SeMIGCN)~\cite{hattori2024learning}.
\item Whole Mesh Generation + Stitch Back: we further compare with public AR mesh generation models (BPT~\cite{weng2025BPT} and MeshAnythingV2~\cite{chen2025meshanything}) using a regenerate-whole-and-stitch-back scheme.
\end{enumerate}
For detailed baseline implementations and more, please refer to \autoref{app:baselines}.

\vspace{-5pt}

\paragraph{Datasets.} Our model is trained on public datasets 
(ShapeNetV2~\cite{chang2015shapenet}, Objaverse~\cite{deitke2023objaverse}, ObjaverseXL~\cite{deitke2023objaverseXL}, etc) converted following
the pipeline of~\cite{liu2026quadgpt}. For testing, we collect 256 low-poly meshes of 
representative categories generated from point clouds of corresponding dense meshes 
using independent AR mesh generation models.
For each of them, we sample 4 random target regions for local editing.

\vspace{-5pt}

\paragraph{Metrics.}
Because local editing requires both seamless boundary attachment and
geometric improvement, standard global mesh metrics are not sufficient on
their own. We therefore use task-specific metrics that measure
boundary correctness and local geometric quality. Specifically, we report
\textbf{A-VMR} (Average Vertex Matching
Ratio), the mean boundary-vertex match rate; \textbf{O-CDIR} (One-way
CD Improvement Rate), the average relative reduction in patch-to-reference one-way Chamfer distance
among perfect-match samples; \textbf{CD-PR} (Chamfer Distance Positive
Rate), the fraction of test samples whose generated patch achieves a
lower one-way Chamfer distance than the original target patch; and
\textbf{\#F-Inc} (Face Increase Ratio),
the relative face-count change.
Formal and complete
definitions are provided in \autoref{app:metrics}.

\subsection{Main Results}
\label{ssec:main-results}

\begin{figure}[t]
    \centering
    \includegraphics[width=\textwidth]{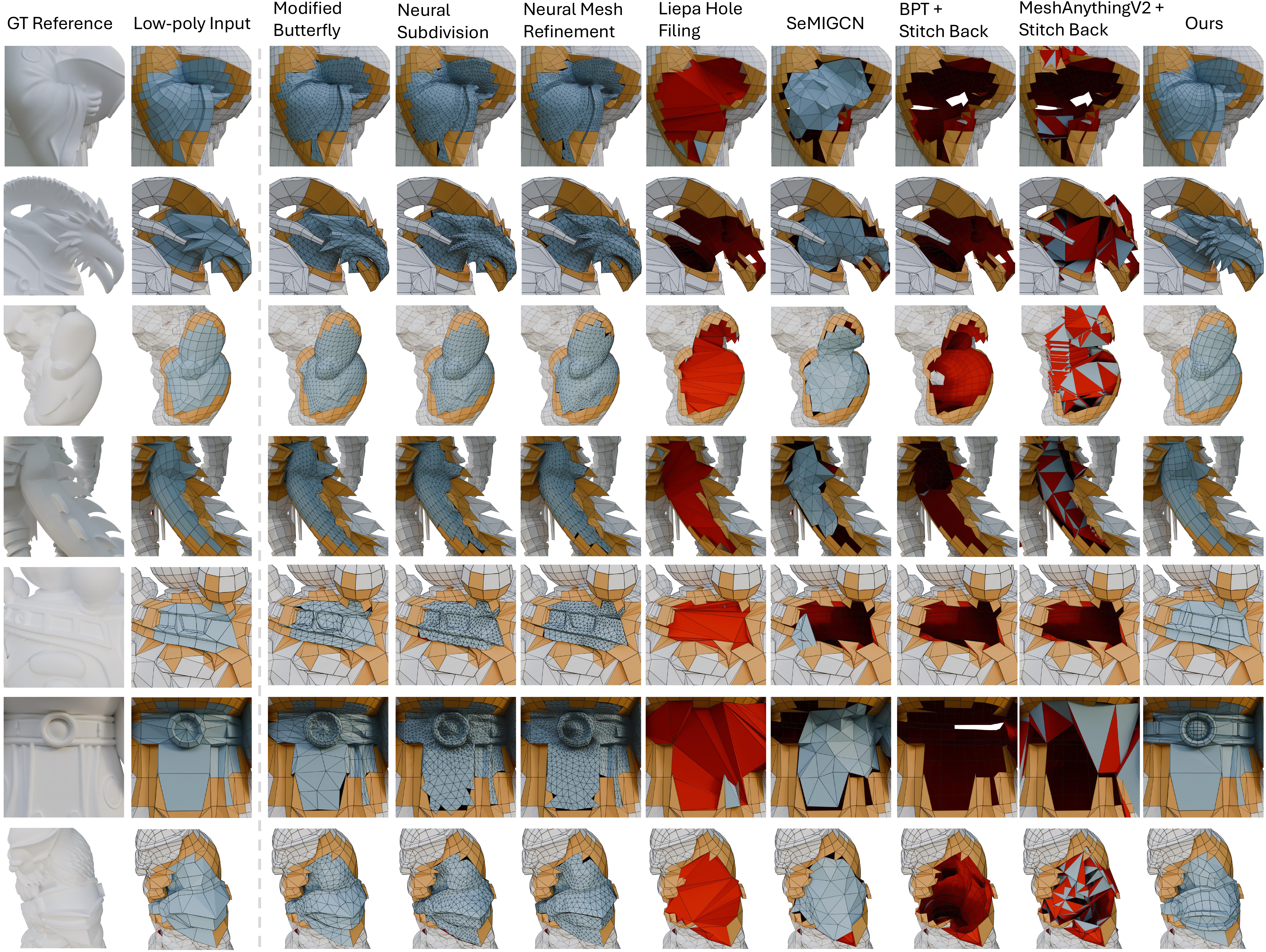}
    \caption{
        \textbf{Qualitative comparisons.}
        Ours is the only method that can efficiently and accurately generate 
        the local low-poly mesh, with seamless stitching 
        and good alignment with nearby edge flows. \textcolor[RGB]{143,33,19}{Dark red} faces indicate back faces (i.e., the opposite side defined by the vertex winding).
    }
    \label{fig:qualitative_baseline_results}
\end{figure}

We show our quantitative results in~\autoref{tab:main-quantitative-results} and
qualitative results in~\autoref{fig:qualitative_baseline_results}. As shown,
our method is able to edit local low-poly mesh efficiently and accurately.
Others are either not local and affect nearby already existing mesh faces
(e.g., Neural Subdivision, Neural Mesh Refinement, BPT, MeshAnythingV2, etc.),
not accurate (unable to refer to the reference geometry), or
not efficient (e.g., BPT and MeshAnythingV2 have to generate the whole low-poly mesh). 
BPT and MeshAnythingV2 often fail to generate the whole mesh successfully, 
resulting in no face to be stitched back. For more explanations on baseline performances and more our results, 
please refer to App. \ref{app:baselines} and App. \ref{app:more-results}. 

\begin{table}[t]
\vspace{-12pt}
\centering
\caption{\textbf{Main results.}
All numerical values presented are percentages (\%).}
\label{tab:main-quantitative-results}
\footnotesize
\setlength{\tabcolsep}{4pt}
\resizebox{\linewidth}{!}{%
\begin{tabular}{@{}lcccccc@{}}
    \toprule
    \textbf{Method}
        & \textbf{A-VMR}$\uparrow$
        & \textbf{O-CDIR}$\uparrow$
        & \textbf{CD-PR}$\uparrow$
        & \textbf{\#F-Inc}
        & \textbf{Locality}
        & \textbf{GT Ref} \\
    \midrule
    \multicolumn{7}{@{}l}{\scriptsize\itshape Refinement} \\
    Modified Butterfly
        & 100.00 & $-2.00$ & 25.00 & 300.00 & \cmark & \xmark \\
    Neural Subdivision
        & 0.00 & $-0.002$ & 38.28 & 852.25 & \xmark & \xmark \\
    Neural Mesh Refinement
        & 33.68 & 0.007 & 60.94 & 845.21 & \xmark & \xmark \\
    \midrule
    \multicolumn{7}{@{}l}{\scriptsize\itshape Hole Filling / Repairing} \\
    Liepa Hole Filling
        & 67.64 & -220.82  & 5.47 & -73.97 & \cmark & \xmark \\
    SeMIGCN
        & 0.00 & -191.62 & 0.93 & -41.29 & \xmark & \xmark \\
    \midrule
    \multicolumn{7}{@{}l}{\scriptsize\itshape Regen Whole + Stitch Back} \\
    BPT + Stitch Back 
        & 2.65 & -67.25 & 0.00 & -84.98 & \xmark & \cmark \\
    MeshAnythingV2 + Stitch Back 
        & 26.90 & -335.28 & 0.00 & -20.45 & \xmark & \cmark \\
    \midrule
    \rowcolor{fullrowblue}
    \textbf{MeshFIM (Ours)}
        & 99.75 & 8.25 & 95.54 & 55.83 & \cmark & \cmark \\
    \bottomrule
\end{tabular}%
}
\end{table}

\subsection{Ablation Studies}
\label{ssec:ablation}

In addition to the metrics above, we report \textbf{OvR} (Overflow
Rate), the fraction of test samples whose overflow ratio exceeds the
threshold $\theta_{\mathrm{ovf}}$; and \textbf{A-Overflow} (Average
Overflow), the mean overflow ratio across all test samples (see
Appendix~\ref{app:metrics} for formal definitions). All ablation rows
are trained for 250k iterations under the same setting.

\begin{table}[t]
\centering
\caption{\textbf{Ablation study.}
Cumulative component ablation; each row adds one component to the
previous configuration.
$\dagger$~Full model (all components enabled).
$\dagger\dagger$~Values in parentheses report the
\emph{no-edit reference lower bound} for OvR and A-Overflow: the values
obtained when the original target faces are used as the generated
patch. The reference lower bounds are non-zero because OvR and A-Overflow metrics use a distance proximity threshold
$\epsilon_{\mathrm{ovf}}$ (Alg.~\ref{alg:gate-merge}).
$\ddagger$~Naive token-axis concatenation replaces the gated fusion.
$\lozenge$~Each encoder branch uses independently sampled query positions
instead of the shared set $\mathbf{q}$.
All values are percentages (\%).}
\label{tab:ablation}
\footnotesize
\setlength{\tabcolsep}{4pt}
\resizebox{\linewidth}{!}{%
\begin{tabular}{@{}lccc l l@{}}
    \toprule
    \textbf{Configuration}
        & \textbf{PMR}$\uparrow$
        & \textbf{A-VMR}$\uparrow$
        & \textbf{O-CDIR}$\uparrow$
        & \textbf{OvR}$\downarrow$
        & \textbf{A-Overflow}$\downarrow$ \\
    \midrule
    Baseline
        & 52.12 & 94.59 & 5.03 & 25.00 & 17.81 \\
    A +Boundary Vertex Markers
        & 69.53 & 98.49 & 6.15 & 14.84 & 15.12 \\
    B +Expanded Context ($w\!=\!3$)
        & 75.78 & 99.12 & 5.49 & 10.16 & 10.52 \\
    C +Context Augmentation
        & 84.38 & 99.18 & 7.12 & 10.16 & 12.13 \\
    D +Context Positional Emb.
        & 85.16 & 99.45 & 7.48 & \phantom{0}9.38 & 10.75 \\
    \rowcolor{fullrowblue}
    E +Low-Poly Geometry Enc. (Full$^\dagger$)
        & \textbf{88.28} & \textbf{99.75} & \textbf{8.25} 
        & \phantom{0}\textbf{6.25}\,{\tiny\color{gray}$(6.25)^{\dagger\dagger}$}
        & \phantom{0}\textbf{4.40}\,{\tiny\color{gray}$(3.49)^{\dagger\dagger}$} \\
    \midrule
    \multicolumn{6}{@{}l}{\scriptsize\itshape Low-poly geo.\ enc.\ ablation} \\
    Full$^\dagger$ w/ naive concat$^\ddagger$
        & 85.16 & 99.51 & 7.50 & \phantom{0}8.59 & \phantom{0}9.27 \\
    Full$^\dagger$ w/ no shared query$^\lozenge$
        & 85.31 & 99.48 & 7.31 & \phantom{0}9.38 & 10.61 \\
    \bottomrule
\end{tabular}%
}
\end{table}

\begin{figure}[t]
    \centering
    \includegraphics[width=\textwidth]{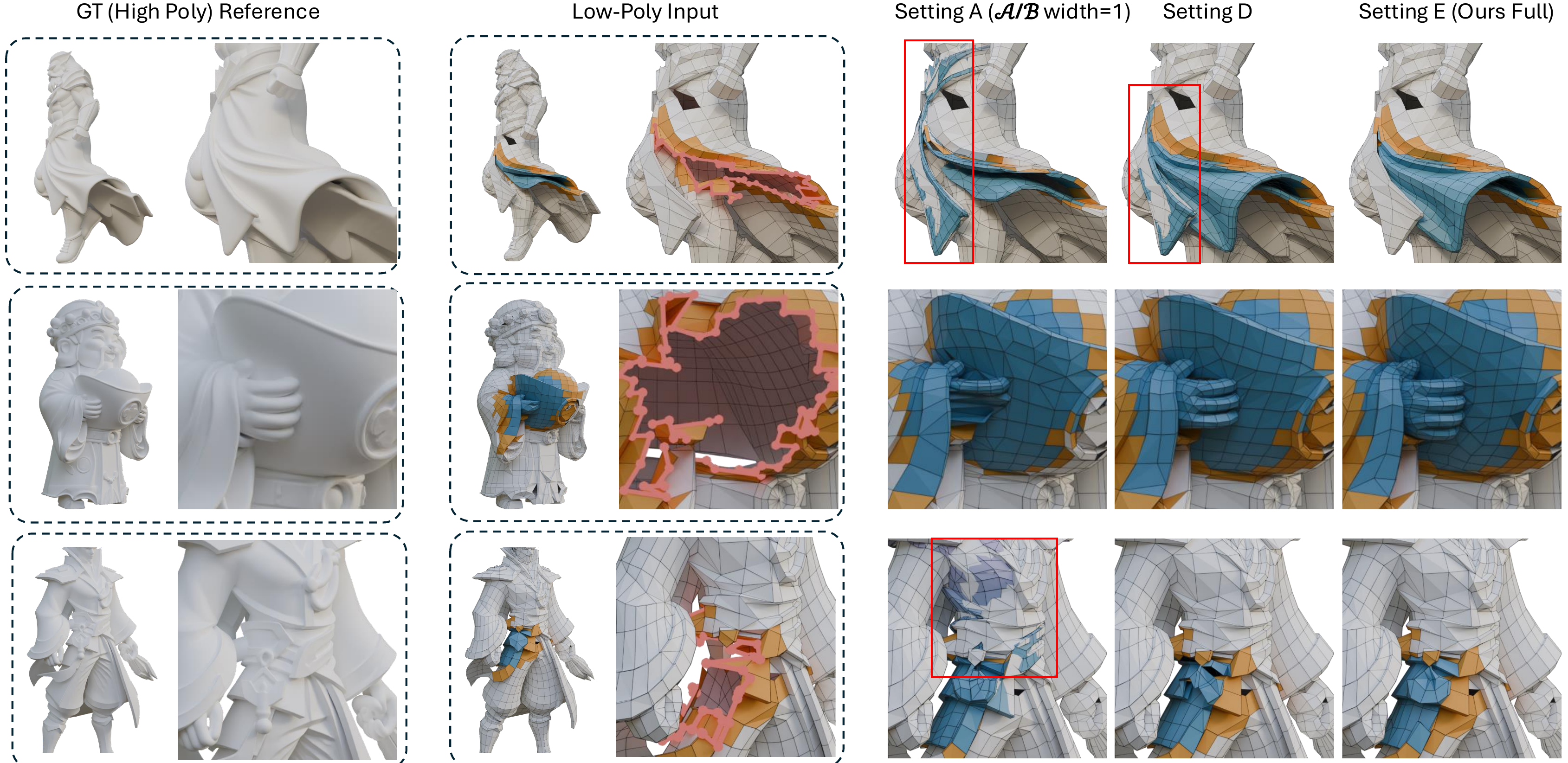}
    \caption{\textbf{Ablation qualitative results.}
    Settings A, D, and E correspond to configurations in \autoref{tab:ablation}.
    \textcolor[RGB]{234,51,35}{Red rectangles} indicate overflow regions.
    Expanding the context to width 3 (A$\to$D) reduces overflow,
    and the low-poly geometry encoder (D$\to$E) further suppresses it
    while improving generation quality.
    }
    \label{fig:overflow}
    \vspace{-10pt}
\end{figure}

\begin{figure}[!htb]
    \centering
    \includegraphics[width=0.95\textwidth]{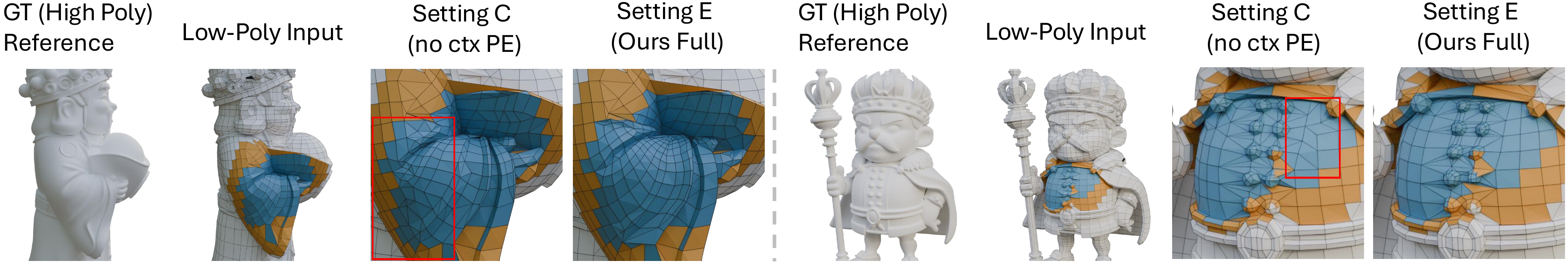}
    \caption{\textbf{Context positional embeddings on
    topology continuity.}  Settings C and E correspond to configurations in \autoref{tab:ablation}.
    \textcolor[RGB]{234,51,35}{Red rectangles} indicate the regions with poor edge flow.
    Context positional embeddings improve the edge flow so that it is aligned with the surrounding mesh.}
    \label{fig:ctx-pos-emb}
\end{figure}

Table~\ref{tab:ablation} reports the cumulative component ablation.
Boundary vertex markers provide the largest single-component gain in PMR (+17.41 pp), confirming
that explicit attachment cues are critical for seamless stitching.
Expanding the context from $w\!=\!1$ to $w\!=\!3$ sharply reduces both overflow metrics (OvR 14.84$\to$10.16, A-Overflow 15.12$\to$10.52), consistent with
Figure~\ref{fig:overflow}: a wider context supplies sufficient geometric evidence to keep generation local.
Context augmentation improves PMR by 8.60 pp, showing that noisy boundaries during training close the train-inference gap.
Context positional embeddings
deliver additional topology quality, matching qualitative trends in
Figure~\ref{fig:ctx-pos-emb} where edge flow becomes aligned with surrounding patterns.
Finally, the low-poly geometry encoder
reduces overflow further (OvR 9.38$\to$6.25, A-Overflow 10.75$\to$4.40) by giving the decoder a
continuous signal of where existing geometry
already covers the reference surface. Our full model approaches
the no-edit reference lower bound (OvR 6.25 vs.\ 6.25, A-Overflow 4.40 vs.\ 3.49), indicating residual overflow stems from the proximity threshold rather than generation errors.

\begin{wrapfigure}{l}{0.44\textwidth}
    \centering
    \vspace{-12pt}
    \includegraphics[width=\linewidth]{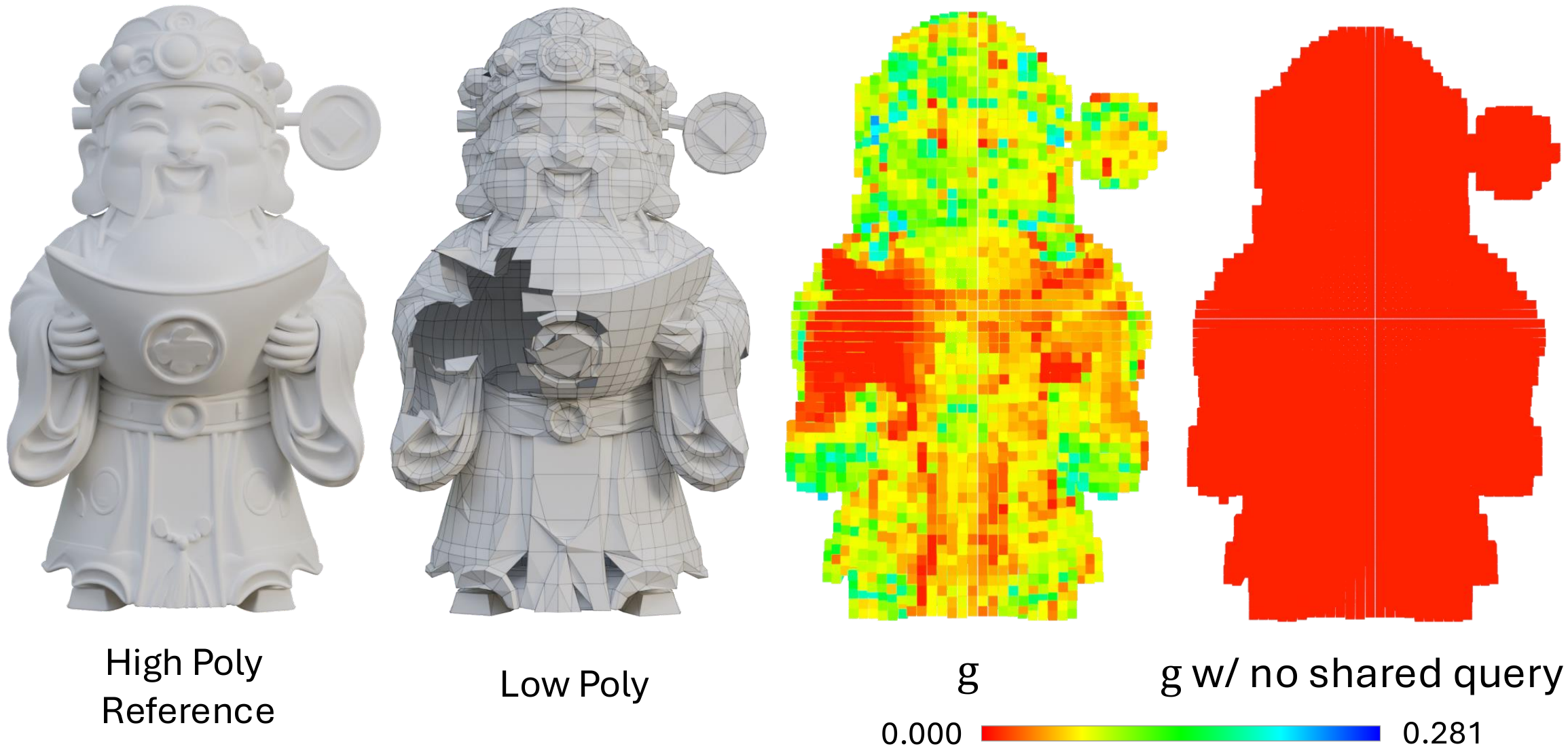}
    \vspace{-14pt}
    \caption{\textbf{Gate-value visualization} (shared \emph{vs.}\ unshared query positions).}
    \label{fig:gate-vis}
    \vspace{-2pt}
\end{wrapfigure}
The ($\ddagger$~naive concat) row replaces the gated design
with a naive token-wise concatenation of the two latent sequences,
$\hat{\mathbf{z}} = \mathbf{z}_{\text{gt}} \oplus \mathbf{z}_{\text{lp}}$
(where $\oplus$ denotes concatenation along the token axis, doubling
the cross-attention context length). Unlike the channel-wise
concatenation used inside the gate in Eq.~\ref{eq:lp-gate}, this naive
variant performs no position-wise alignment and no explicit gating.
The ($\lozenge$~no shared query) row further demonstrates the importance of
position-wise alignment: each encoder branch uses independently
sampled query positions, so the $i$-th tokens
$\mathbf{z}_{\text{gt},i}$ and $\mathbf{z}_{\text{lp},i}$ correspond
to different spatial locations. Without positional alignment, the gate
receives uninformative input and collapses to near-zero values across
all positions (Figure~\ref{fig:gate-vis}).

%
%


\FloatBarrier

\vspace{-4pt}
\section{Applications}
\label{sec:applications}
\vspace{-2pt}

We demonstrate two applications of MeshFIM: interactive brush-based
editing (section. \ref{ssec:brush}) and automatic defect
repair~(section. \ref{ssec:hole-repair}). Results are shown in the teaser
(\autoref{fig:teaser}).

\vspace{-3pt}
\subsection{Interactive Brush-Based Target Selection}
\label{ssec:brush}
\vspace{-2pt}

We provide a web-based 3D interface where the user paints faces with
a radius-adjustable brush to define $\target$. The brush selects
faces under the cursor; each stroke is reduced to its largest
connected component on the face adjacency graph via BFS. Subsequent
strokes that are adjacent to the current selection are merged
automatically, ensuring that $\target$ always remains a single
connected region. See Appendix~\ref{app:brush-interface} for
interface and implementation details as well as a video record of usage in supp.

\vspace{-3pt}
\subsection{Automatic Defect Detection}
\label{ssec:hole-repair}
\vspace{-2pt}

\begin{figure}[!h]
    \centering
    \includegraphics[width=\textwidth]{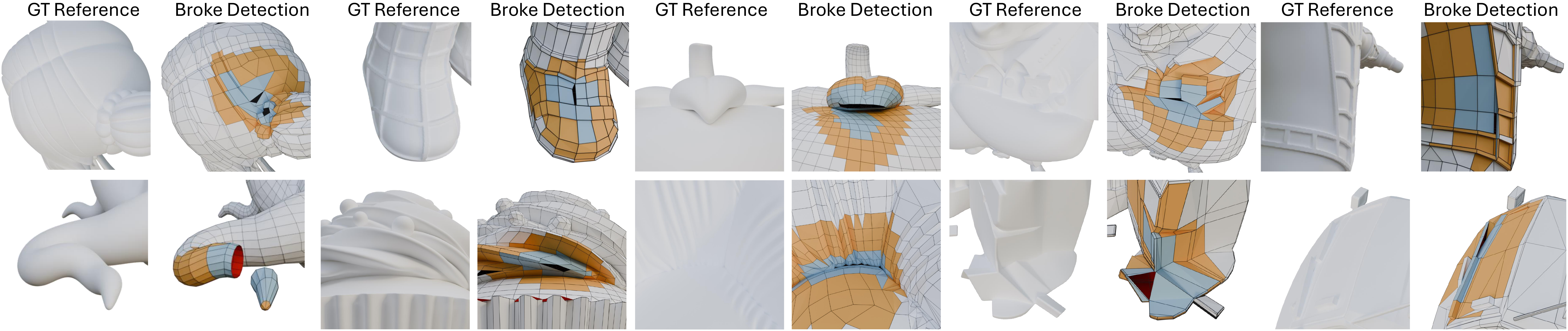}
    \caption{\textbf{Broken detection examples.}
    Our detection algorithm (Alg.~\ref{alg:detect} + Alg.~\ref{alg:region}) reliably identifies
    defects across a wide range of severity, from large missing-face
    clusters to subtle single-face flips or micro-holes spanning
    only a few pixels, demonstrating robustness to diverse defect types.}
    \label{fig:detection-examples}
\end{figure}

AR mesh generators can produce defective local regions such as
missing face clusters or malformed topology. We build an automatic
repair pipeline that can detect generation regions (see~\autoref{fig:detection-examples}) via 
carefully designed graphics heuristics.
Repair can be done automatically as well by involving 
detecting, repair (MeshFIM inference), quality gating and merging. 
This whole pipeline can be run iteratively.
Full implementation details and more results on this are in Appendix~\ref{app:hole-repair}.

\vspace{-3pt}

\vspace{-2pt}

\section{Conclusion, limitations, and future work}
\label{sec:conclusion}

We presented MeshFIM, the first Fill-in-the-Middle framework for autoregressive
low-poly mesh generation, enabling local editing without full regeneration.
Five complementary designs jointly enforce boundary attachment, suppress overflow,
and focus generation on the missing region. 
Experiments show our unique and significant advantages over a broad range of baselines in 
the sense of  generated mesh topology quality and seamless alignment with surrounding 
existing mesh faces. One limitation of our current method is that it is built on the assumption 
that the target region $\target$ is a single connected region. This works for most of the cases 
but may need to be relaxed when it comes to broken regions that involve multiple individual and 
disconnected components, which we leave to future work.


\clearpage
\bibliographystyle{plainnat}
\bibliography{references}

\newpage
\appendix
\section{Detailed Metric Definitions}
\label{app:metrics}

We provide formal definitions of the evaluation metrics used in
the main paper.
Let $S=\{s_1,\dots,s_N\}$ denote the test set of $N$ samples.
For each sample $s$, we have the ground-truth mesh $G^{(s)}$, the
context region $(\context)^{(s)}$, the target region
$\target^{(s)}$, and the generated replacement $(\generated)^{(s)}$.
Let $\tau$ denote the vertex-matching tolerance (set to $10^{-6}$
throughout).

\paragraph{Matched boundary vertices.}
For sample $s$, let $V_{\partial}^{(s)}$ denote the exposed
boundary-vertex set. The matched boundary vertices are those boundary 
vertices that coincide almost exactly (within $\tau$) with at least one
generated vertex:
\begin{equation}
    V_{\text{M}}^{(s)}
    = \bigl\{\,v \in V_{\partial}^{(s)}
      : \min_{u \in V_{\generated}^{(s)}} \|v - u\|_2 < \tau
      \bigr\}.
\end{equation}

\paragraph{Vertex matching ratio.}
The per-sample vertex matching ratio measures how many boundary vertices are reproduced within the generated region:
\begin{equation}
       r(s) = \frac{|V_{\text{M}}^{(s)}|}{|V_{\partial}^{(s)}|}
\end{equation}

\paragraph{PMR (Perfect Matching Ratio).}
The fraction of samples achieving perfect matching of all boundary vertices:
\begin{equation}
    \mathrm{PMR}
    = \frac{1}{N}\sum_{s \in S} \mathbf{1}\!\bigl[r(s) = 1\bigr].
\end{equation}

\paragraph{A-VMR (Average Vertex Matching Ratio).}
The mean vertex matching ratio across all samples:
\begin{equation}
    \mathrm{A\text{-}VMR}
    = \frac{1}{N}\sum_{s \in S} r(s).
\end{equation}

\paragraph{One-way Chamfer distance.}
For a point set $X$ sampled from mesh $A$ and a point set $Y$
sampled from mesh $B$:
\begin{equation}
    \mathrm{CD}(A \to B)
    = \frac{1}{|X|}\sum_{x \in X}
      \min_{y \in Y} \|x - y\|_2.
\end{equation}
We sample $100{,}000$ points from the ground truth $G$ and
$20{,}000$ points from $\target$ and $\generated$.

The following three metrics are computed only over the
\emph{perfect-match subset}
$S_P = \{s \in S : r(s) = 1\}$. We use a one-way distance because
$\target^{(s)}$ and $\generated^{(s)}$ are local patches, whereas
$G^{(s)}$ is the full mesh. Accordingly, $\mathrm{CD}(A \to G^{(s)})$
checks whether the local patch $A$ lies on the reference surface
without penalizing the large part of $G^{(s)}$ outside the edited
region.

\paragraph{O-CDIR (One-way CD Improvement Rate).}
The average relative reduction in one-way Chamfer distance:
\begin{equation}
    \mathrm{O\text{-}CDIR}
    = \frac{1}{|S_P|}\sum_{s \in S_P}
      \frac{
        \mathrm{CD}(\target^{(s)} \!\to\! G^{(s)})
        - \mathrm{CD}(\generated^{(s)} \!\to\! G^{(s)})
      }{
        \mathrm{CD}(\target^{(s)} \!\to\! G^{(s)})
      }.
\end{equation}

\paragraph{\#F-Inc (Face Increase Ratio).}
The average relative change in face count:
\begin{equation}
    \text{\#F-Inc}
    = \frac{1}{|S_P|}\sum_{s \in S_P}
      \frac{|F_{\generated}^{(s)}| - |F_{\target}^{(s)}|}
           {|F_{\target}^{(s)}|}.
\end{equation}

\paragraph{CD-PR (Chamfer Distance Positive Rate).}
The fraction of all test samples for which the generated patch achieves
a lower one-way Chamfer distance to the reference than the original
target patch:
\begin{equation}
    \mathrm{CD\text{-}PR}
    = \frac{1}{N}\sum_{s \in S}
      \mathbf{1}\!\bigl[
        \mathrm{CD}(\generated^{(s)} \!\to\! G^{(s)})
        < \mathrm{CD}(\target^{(s)} \!\to\! G^{(s)})
      \bigr].
\end{equation}

\paragraph{OvR (Overflow Rate).}
Let $\theta_{\mathrm{ovf}}$ be the overflow ratio threshold
(set to $0.01$ throughout; see Alg.~\ref{alg:gate-merge}).
OvR counts the fraction of test samples whose per-sample overflow ratio
$O(s)$ exceeds $\theta_{\mathrm{ovf}}$:
\begin{equation}
    \mathrm{OvR}
    = \frac{\bigl|\{s \in S : O(s) > \theta_{\mathrm{ovf}}\}\bigr|}{N}.
\end{equation}

\paragraph{A-Overflow (Average Overflow Percentage).}
We define the per-sample overflow ratio $O(s)$ as the fraction of
points sampled from $\generated^{(s)}$ that lie within distance
$\epsilon_{\mathrm{ovf}}$ of the residual faces of the original mesh (excluding
$\target$ and $\context$).
Let $S_{>\theta_{\mathrm{ovf}}} = \{s \in S : O(s) > \theta_{\mathrm{ovf}}\}$ be the subset of samples
with non-negligible overflow.
The average overflow percentage is:
\begin{equation}
    \text{A-Overflow}
    = \frac{1}{|S_{>\theta_{\mathrm{ovf}}}|}\sum_{s \in S_{>\theta_{\mathrm{ovf}}}} O(s).
\end{equation}
We set $\epsilon_{\mathrm{ovf}}$ to the default overflow detection threshold
throughout all experiments.

\paragraph{No-edit reference lower bound for OvR and A-Overflow.}
Because $O(s)$ counts points within distance $\epsilon_{\mathrm{ovf}}$
of the residual faces, samples from $\generated^{(s)}$ near the shared
boundary are always close to the neighboring context and residual
geometry. Consequently, OvR and A-Overflow are bounded below by a
nonzero residual induced by the proximity threshold, independent of
the generator. We estimate this residual by the \emph{no-edit
reference}: substituting the original target faces for the generated
patch ($\generated^{(s)} \equiv \target^{(s)}$) and evaluating the
same two metrics. On our ablation test set this yields
$\mathrm{OvR}=6.25\%$ and $\mathrm{A\text{-}Overflow}=3.49\%$, which
we report in parentheses under the Full row of
Table~\ref{tab:ablation}. No method can go below these values; lower
reported numbers would indicate an evaluation artifact rather than a
genuinely overflow-free output.

\section{Baseline Implementation Details}
\label{app:baselines}

Because our test meshes contain open-boundary low-poly meshes
(with quads triangulated for baselines that require triangles),
several baselines need non-trivial preprocessing or post-processing
to produce a local patch for evaluation.
Table~\ref{tab:baseline-prereqs} summarizes the key requirements;
the rest of this section details only the non-obvious steps.

\begin{table}[h]
\centering
\caption{Baseline preprocessing and evaluation setup.
``Closed mesh'' means the method requires a watertight manifold
input. ``Global'' means the method operates on the entire mesh
(not just the target region).}
\label{tab:baseline-prereqs}
\footnotesize
\resizebox{\textwidth}{!}{%
\begin{tabular}{@{}lcccl@{}}
    \toprule
    \textbf{Method} & \textbf{Closed} & \textbf{Global} & \textbf{Stitch} & \textbf{Preprocessing / Notes} \\
    \midrule
    Modified Butterfly   & \xmark & \cmark & \xmark & Face-id tracking after subdivision \\
    Neural Subdivision   & \cmark & \cmark & \xmark & fTetWild~\cite{hu2020fast} to close mesh + face-id remap \\
    Neural Mesh Refinement  & \cmark & \cmark & \xmark & fTetWild~\cite{hu2020fast} to close mesh + face-id remap \\
    \midrule
    Liepa Hole Filling   & \xmark & --- & \xmark & Remove $\target$; fill boundary loop via min-weight triangulation \\
    SeMIGCN              & \xmark & --- & \xmark & MeshFix to close hole + per-shape GCN optimization (100 iter) \\
    \midrule
    BPT + Stitch Back        & \xmark & \cmark & \cmark & Regenerate whole mesh from point cloud; spatial crop + boundary snap \\
    MeshAnythingV2 + Stitch  & \xmark & \cmark & \cmark & Regenerate whole mesh from point cloud; spatial crop + boundary snap \\
    \midrule
    \textbf{MeshFIM (Ours)}  & \xmark & \xmark & \xmark & Direct local generation; no pre/post-processing needed \\
    \bottomrule
\end{tabular}%
}
\end{table}

\paragraph{Why subdivision baselines are non-local.}
Neural Subdivision~\cite{liu2020neural} and Neural Mesh
Refinement~\cite{zhu2025neural} are \emph{global} subdivision
operators: they take the full mesh connectivity as input and
predict new vertex positions for \emph{every} subdivided face
simultaneously (the network's receptive field spans the entire
mesh via pooling hierarchies). There is no mechanism to subdivide
only a selected region while keeping the rest unchanged.

\paragraph{Patch extraction for subdivision baselines.}
Because subdivision is applied globally, we track which original
face each subdivided face descends from: after $L$ levels of
1-to-4 subdivision, each original face produces $4^L$ child faces
that inherit its source id. The target-region patch is then
extracted by collecting all child faces whose inherited source id
belongs to $\target$.
No stitch-back post-processing is applied, because the extracted
patch is not independently generated---it is carved out of the
same subdivided mesh that contains the context.
However, the subdivision \emph{itself} modifies the surrounding
geometry: vertex positions along and near the boundary are
repositioned by the learned (or interpolatory) subdivision rules,
and the boundary of the extracted patch no longer coincides with
the original context boundary vertices. This explains the low
A-VMR scores in Table~\ref{tab:main-quantitative-results}
(0\% for Neural Subdivision, 33.68\% for Neural Mesh Refinement).

\paragraph{fTetWild preprocessing (Neural Subdivision, Neural Mesh Refinement).}
Both methods further assume a closed manifold input and crash on
open-boundary meshes. We close each test mesh using
fTetWild~\cite{hu2020fast} with its \texttt{--manifold-surface}
flag and default envelope ($10^{-3}$ of the bounding-box diagonal).
Because fTetWild may re-triangulate or add cap faces, we remap each
output face to its nearest original face (centroid-to-surface query)
and propagate the original face source id; cap faces (distance
above $10\times$ envelope) are marked with id $-1$.
After subdivision, the target-region patch is extracted by
collecting all output faces whose inherited source id belongs to
$\target$.

\paragraph{Stitch-back scheme (BPT, MeshAnythingV2).}
These models regenerate the \emph{entire} mesh from the
ground-truth point cloud and have no mechanism for local editing.
To extract a local patch for evaluation:
\begin{enumerate}[nosep,leftmargin=*]
    \item \textbf{Spatial crop:} select faces of the regenerated mesh
          whose centroids fall inside the bounding box of the
          original $\target$ (expanded by 5\% of the diagonal).
    \item \textbf{Boundary snap:} for each context boundary vertex,
          find the nearest cropped-patch boundary vertex via
          KD-tree; if within $2\times$ the patch's average edge
          length, snap the patch vertex to the exact context
          position. Unmatched context vertices are appended as
          isolated vertices so the boundary-matching metric can
          still be evaluated.
\end{enumerate}
This generous tolerance favors the baselines; nonetheless,
boundary matching remains poor because the independently generated
mesh has no constraint to align with the surrounding context.

\paragraph{SeMIGCN per-shape optimization.}
Unlike methods that use a pre-trained model, SeMIGCN trains a
small GCN from scratch for every test sample (100 iterations).
The target faces are removed to expose a hole, MeshFix closes the
mesh conservatively, and the GCN learns vertex displacements
in a self-supervised loop. A Laplacian-based refinement step
then smooths the result. Because the optimization is
unsupervised and shape-agnostic, it relies on local smoothness
priors rather than learned geometry, explaining its limited
geometric accuracy.

\section{Automatic Defect Repair}
\label{app:hole-repair}

\subsection{Algorithmic Details}
\begin{figure}[ht]
\includegraphics[width=\textwidth]{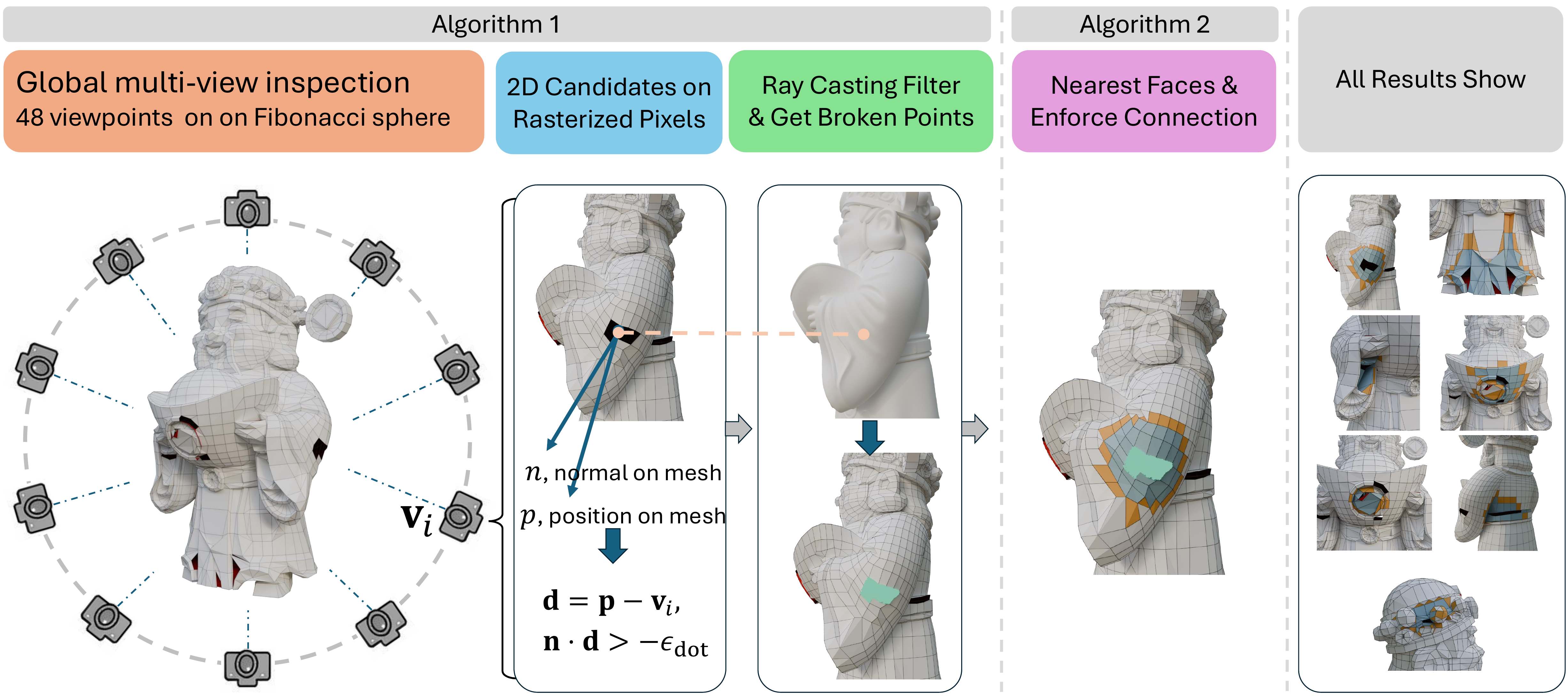}
\caption{The automatic defect detection pipeline. 
Our pipeline provides accurate defect detection to various cases via 
delicately designed graphics heuristics.}
\label{fig:algorithm_pipeline}
\end{figure}

We describe the four stages of the automatic hole-repair
pipeline introduced in \autoref{ssec:hole-repair}: 
(1) candidate ROIs in space are found by finding broken points Alg~\autoref{alg:detect}. 
(2) based on broken points detected, we extract related mesh faces with 
topology connection enforced Alg~\autoref{alg:region}. 
(3) MeshFIM is run based on the detection information.
(4) we filter and merge the local generation based on metrics, 
including the overflow metrics mentioned before Alg~\autoref{alg:gate-merge}.
Step (1) to (4) can be looped to iteratively repair the mesh Alg~\autoref{alg:iter-loop}.
All default hyper-parameters are listed in each algorithm caption. 
\autoref{fig:algorithm_pipeline} shows the overall pipeline of defect detection 
(Alg~\autoref{alg:detect} and Alg~\autoref{alg:region}).

\subsection{Repair Results}

We evaluate the iterative repair pipeline on 2560 meshes generated by
an independent AR mesh generator. Table~\ref{tab:repair-stats} reports
the overall repair statistics and the per-round breakdown.

\begin{table}[h]
\centering
\caption{\textbf{Iterative repair statistics} on 2560 generated meshes
(max $T\!=\!4$ rounds).}
\label{tab:repair-stats}
\footnotesize
\begin{tabular}{@{}lrc@{}}
    \toprule
    \textbf{Category} & \textbf{Count} & \textbf{Percentage} \\
    \midrule
    Total objects & 2560 & --- \\
    Initially damaged & 2417 & 94.4\% \\
    Never damaged (no defects detected) & 143 & 5.6\% \\
    \midrule
    Successfully repaired & 2149 & 88.9\% of damaged \\
    Still broken after 4 rounds & 268 & 11.1\% of damaged \\
    \bottomrule
\end{tabular}
\vspace{6pt}

\begin{tabular}{@{}lccc@{}}
    \toprule
    \textbf{Round} & \textbf{Newly Fixed} & \textbf{Cumulative Fixed} & \textbf{Cumulative \%} \\
    \midrule
    1 & 1573 & 1573 & 65.1\% \\
    2 & 307 & 1880 & 77.8\% \\
    3 & 168 & 2048 & 84.7\% \\
    4 & 101 & 2149 & 88.9\% \\
    \bottomrule
\end{tabular}
\end{table}

Out of 2417 damaged meshes, 1573 (65.1\%) are fully repaired after a single
round. Subsequent rounds fix progressively fewer objects as the remaining
defects tend to be more severe or involve larger disconnected regions.
After 4 rounds the pipeline achieves an overall repair rate of 88.9\%.

A mesh is marked as ``successfully repaired'' when Alg.~\ref{alg:detect}
detects zero broken-point candidates in the subsequent round.
Note that our detection algorithm is deliberately sensitive: it renders the
mesh from 48 viewpoints at high resolution ($640\!\times\!640$) and confirms
every candidate against the reference mesh via ray casting, so even subtle
artifacts such as a single flipped face or a tiny hole spanning only a few
pixels are reliably flagged (see Figure~\ref{fig:detection-examples} for
examples of detected defects at various severity levels).
The repair criterion is therefore strict---a
mesh passes only when no defect survives this dense multi-view inspection.

\begin{algorithm}[h]
\caption{Broken Point Detection.
Default parameters: $N_v\!=\!48$ views, rasterization resolution
$H\!=\!640$, dot-product threshold $\epsilon_{\mathrm{dot}}\!=\!10^{-4}$,
DBSCAN~\citep{ester1996density} radius $\epsilon_{\mathrm{cls}}\!=\!0.05$,
minimum cluster size $\tau_{\mathrm{bp}}\!=\!10$.}
\label{alg:detect}
\begin{algorithmic}[1]
\REQUIRE Mesh to repair $\mesh$, reference mesh $\mesh_{\mathrm{ref}}$
\ENSURE  Broken-point clusters $\{C_1,\dots,C_K\}$
\STATE Normalize $\mesh$ and $\mesh_{\mathrm{ref}}$ to unit sphere.
\STATE Generate $N_v$ viewpoints on a Fibonacci sphere.
\FOR{each viewpoint $\mathbf{v}_i$}
  \STATE Rasterize $\mesh$ at resolution $H$; record per-pixel face
         normal $\mathbf{n}$ and position $\mathbf{p}$.
  \STATE Compute view direction
         $\mathbf{d} = \mathbf{p} - \mathbf{v}_i$.
  \STATE Mark pixel as \emph{candidate} if
         $\mathbf{n} \cdot \mathbf{d} > -\epsilon_{\mathrm{dot}}$
         (back-face test).
  \STATE For each candidate, cast a ray from $\mathbf{v}_i$ through
         $\mathbf{p}$ against $\mesh_{\mathrm{ref}}$; confirm the
         candidate if the ray hits a correctly oriented reference face
         first.
\ENDFOR
\STATE Cluster all confirmed broken points with DBSCAN
       ($\epsilon_{\mathrm{cls}}$); discard clusters with fewer than
       $\tau_{\mathrm{bp}}$ points.
\RETURN $\{C_1,\dots,C_K\}$
\end{algorithmic}
\end{algorithm}

\begin{algorithm}[h]
\caption{Damage Region Extraction.
Default parameters: B-expansion width $w_{\target}\!=\!3$,
context-expansion width $w_{\context}\!=\!1$.}
\label{alg:region}
\begin{algorithmic}[1]
\REQUIRE Broken-point clusters $\{C_k\}$, mesh $\mesh$
\ENSURE  Damage groups $\{(\target_j,\;(\context)_j)\}$
\FOR{each cluster $C_k$}
  \STATE Map every point in $C_k$ to its nearest face in $\mesh$
         $\rightarrow$ seed faces $F_{\mathrm{seed}}^{(k)}$.
  \STATE If seed faces are not face-connected, link disjoint components
         via shortest-path BFS on the face adjacency graph.
  \STATE BFS-expand $F_{\mathrm{seed}}^{(k)}$ by $w_{\target}$ rings
         $\rightarrow$ target region $\target_k$.
  \STATE BFS-expand the boundary of $\target_k$ by $w_{\context}$ ring
         $\rightarrow$ context region $(\context)_k$.
\ENDFOR
\STATE Merge groups whose $\target$ or $\context$ regions overlap
       using Union-Find.
\RETURN $\{(\target_j,\;(\context)_j)\}$
\end{algorithmic}
\end{algorithm}

\begin{algorithm}[h]
\caption{Quality Gate \& Merge.
Default parameters: overflow distance
$\epsilon_{\mathrm{ovf}}\!=\!1/256$, overflow ratio threshold
$\theta_{\mathrm{ovf}}\!=\!0.01$, weld tolerance
$\tau_{\mathrm{weld}}\!=\!10^{-6}$.}
\label{alg:gate-merge}
\begin{algorithmic}[1]
\REQUIRE Current mesh $\mesh_{i-1}$, target region $\target$,
         generated patch $\generated$
\ENSURE  Updated mesh $\mesh_i$ or \textsc{reject}
\STATE \textbf{// Quality Gate}
\STATE Sample point cloud $\pointcloud_{\generated}$ on $\generated$.
\STATE Let $K = \mesh_{i-1} \setminus (\target \cup (\context))$
       be the remaining mesh.
\STATE Compute overflow ratio:
       fraction of $\pointcloud_{\generated}$ within distance
       $\epsilon_{\mathrm{ovf}}$ of $K$.
\IF{overflow ratio $> \theta_{\mathrm{ovf}}$}
  \RETURN \textsc{reject}
\ENDIF
\STATE \textbf{// Merge}
\STATE Remove target faces:
       $\mesh_i \leftarrow \mesh_{i-1} \setminus \target$.
\STATE Build KD-tree on boundary vertices of $\mesh_i$.
\STATE For each vertex of $\generated$ within
       $\tau_{\mathrm{weld}}$ of a boundary vertex, weld
       (snap) to the existing vertex.
\STATE Append $\generated$ faces and welded vertices to $\mesh_i$.
\RETURN $\mesh_i$
\end{algorithmic}
\end{algorithm}

\begin{algorithm}[h]
\caption{Iterative Hole Repair.
Default parameters: max rounds $T\!=\!4$, false-positive threshold
$\tau_{\mathrm{fp}}\!=\!2$ points.}
\label{alg:iter-loop}
\begin{algorithmic}[1]
\REQUIRE Initial mesh $\mesh_0$, reference mesh $\mesh_{\mathrm{ref}}$
\ENSURE  Repaired mesh $\mesh_T$
\FOR{$i = 1$ \TO $T$}
  \STATE $\{C_k\} \leftarrow$
         \textsc{DetectBrokenPoints}($\mesh_{i-1}$,
         $\mesh_{\mathrm{ref}}$)
         \hfill\COMMENT{Alg.\,\ref{alg:detect}}
  \IF{$\sum_k |C_k| = 0$}
    \RETURN $\mesh_{i-1}$ \hfill\COMMENT{no damage}
  \ELSIF{$\sum_k |C_k| \leq \tau_{\mathrm{fp}}$}
    \RETURN $\mesh_{i-1}$ \hfill\COMMENT{likely false positive}
  \ENDIF
  \STATE $\{(\target_j, (\context)_j)\} \leftarrow$
         \textsc{ExtractRegions}($\{C_k\}$, $\mesh_{i-1}$)
         \hfill\COMMENT{Alg.\,\ref{alg:region}}
  \STATE $\mesh_i \leftarrow \mesh_{i-1}$
  \FOR{each group $(\target_j, (\context)_j)$}
    \STATE $\generated_j \leftarrow$
           \textsc{MeshFIM}($(\context)_j$)
    \STATE $\mesh_i \leftarrow$
           \textsc{QualityGateMerge}($\mesh_i$, $\target_j$,
           $\generated_j$)
           \hfill\COMMENT{Alg.\,\ref{alg:gate-merge}}
  \ENDFOR
\ENDFOR
\RETURN $\mesh_T$
\end{algorithmic}
\end{algorithm}

\section{Percolation-Based Region Sampling}
\label{app:percolation}

To prevent the model from overfitting to regular circular patches, we
sample a fraction of training target regions via stochastic growth on
the face adjacency graph. Starting from a random seed face, the region
expands along the current frontier: for each candidate neighboring
face, we add it with probability~$p$ until the target face budget is
reached. $p$ is sampled uniformly from $[0.55, 0.85]$ per example.
If no frontier face is accepted in one round, the algorithm falls back
to expanding through the remaining frontier so that the sampled region
stays connected and continues to grow. Compared with fixed-radius BFS
patches, this produces connected but irregular regions that better
resemble brush selections or automatically detected defects.

\section{Low Poly Encoder Gate Network Implementation}
\label{app:gate}

This appendix details the gate network $N_{\mathrm{gate}}$ of
section \ref{sec:method} (Eq.~\ref{eq:lp-gate}).

\paragraph{Inputs and shared query.}
The two point-cloud branches share a single Perceiver-IO encoder $E$.
Its cross-attention queries are positions
$\mathbf{q}=\{\mathbf{q}_i\}_{i=1}^{M}$ in Euclidean space obtained
from the reference branch; the low-poly branch reuses the same
$\mathbf{q}$, which is what makes the two latent sequences
$\mathbf{z}_{\text{gt}}, \mathbf{z}_{\text{lp}}\in\mathbb{R}^{M\times d}$
position-aligned.

\paragraph{Architecture of $N_{\mathrm{gate}}$.}
The gate vector $\mathbf{g}\in(0,1)^M$ is produced by a lightweight stack operating
on the channel-wise concatenation of the two latents:
\begin{align*}
    \mathbf{h} &= [\mathrm{LN}(\text{sg}(\mathbf{z}_{\text{gt}}))\,;\;
                     \mathrm{LN}(\text{sg}(\mathbf{z}_{\text{lp}}))]
                 \in \mathbb{R}^{M \times 2d}, \\
    \mathbf{h}   &\leftarrow \mathbf{h} + \mathrm{SelfAttn}(\mathbf{h}), \\
    \mathbf{h}   &\leftarrow \mathbf{h} + \mathrm{FFN}(\mathrm{LN}(\mathbf{h})), \\
    \mathbf{h}   &\leftarrow \mathrm{TransformerEncoder}_{1}(\mathbf{h}), \\
    \mathbf{g}   & = \sigma\bigl(\mathbf{w}^{\top}\mathbf{h} + b\bigr),
\end{align*}
where $\text{sg}(\cdot)$ denotes stop-gradient. The output head is a
linear map to a scalar at each token position; its weights are initialized with
$\mathcal{N}(0, 10^{-3})$ and its bias with $b = -4$, so that at the
start of training $g_i \approx \sigma(-4) \approx 0.018$, yielding
$\hat{\mathbf{z}}\approx\mathbf{z}_{\text{gt}}$ initially. The FFN follows the
standard two-layer GELU design with hidden size $2\cdot 2d$.



\section{Interactive Brush Selection Interface}
\label{app:brush-interface}

\begin{figure}[!th]
    \centering
    \includegraphics[width=\textwidth]{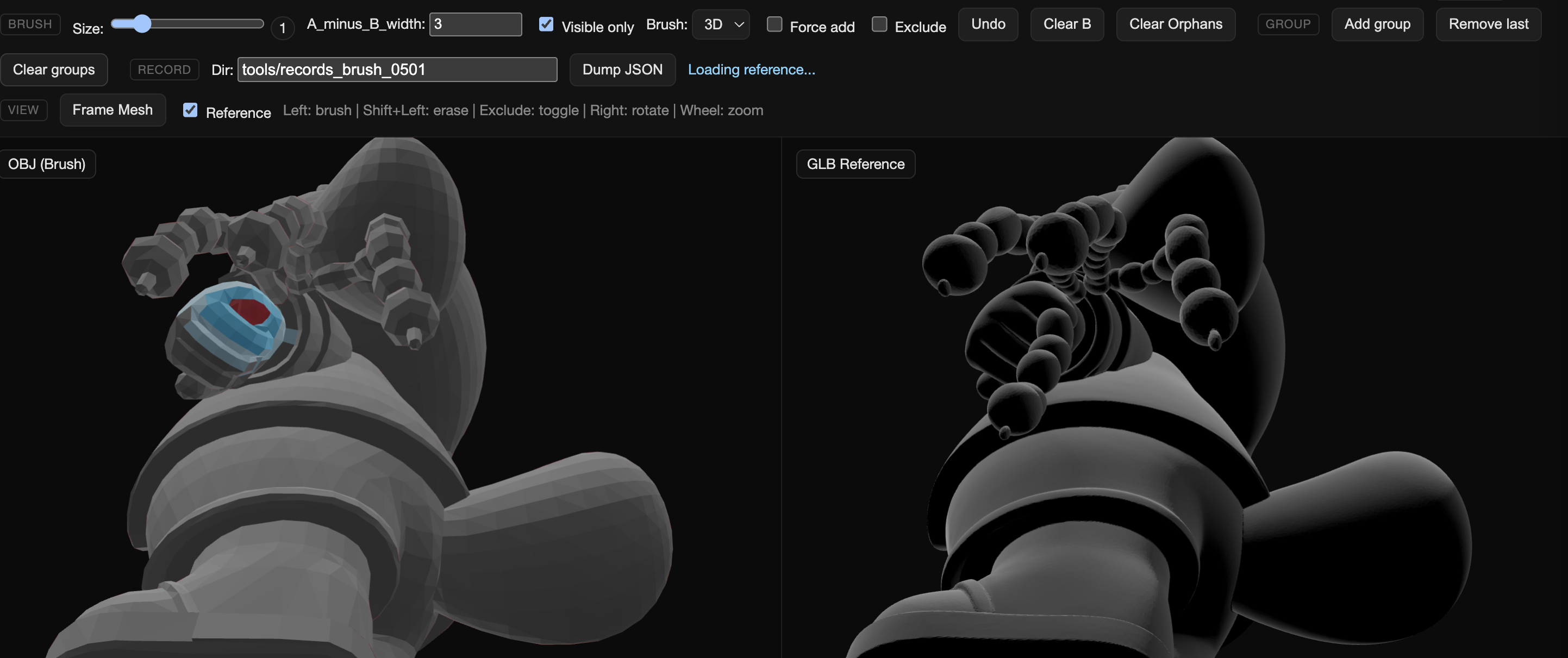}
    \caption{\textbf{Brush editing interface.}
    Left: low-poly mesh with target $\target$
    (\textcolor[RGB]{38,166,255}{blue}) and context $\context$
    (\textcolor[RGB]{115,204,255}{light blue}).
    Back-facing faces are tinted
    \textcolor[RGB]{255,59,59}{dark red} to make it easier for users 
    to observe broken regions.
    Right: reference GT mesh. Top bar: brush size, mode (2D/3D),
    context width, and navigation controls.}
    \label{fig:brush-ui}
\end{figure}

\autoref{fig:brush-ui} shows our browser-based brush selection tool. The
interface presents a synchronized dual-panel view: the left panel
displays the low-poly mesh for brush painting, while the right panel
shows the reference geometry for visual guidance. Both share the same
camera.

\paragraph{Face-level brush.}
The brush selects whole faces via raycasting, supporting two modes:
(i)~a \emph{2D} mode that BFS-expands from the hit face by a
user-specified ring count on the adjacency graph, and (ii)~a \emph{3D}
mode that selects all faces whose centroids fall within a world-space
sphere. An optional backface test restricts selection to visible
geometry.

\paragraph{Connectivity enforcement.}
Each committed stroke is filtered to its largest connected component.
Strokes not adjacent to the existing selection are flagged as orphans
and excluded, guaranteeing that the final $\target$ is always a
connected patch---a requirement of the FIM serialization.

\paragraph{Context preview.}
The tool displays the context ring $\context$ in real time by
BFS-expanding from the boundary of $\target$ with a user-adjustable
width, giving immediate feedback on what the model will condition on.

\section{More results}
\label{app:more-results}

\begin{figure}[!h]
\includegraphics[width=\textwidth]{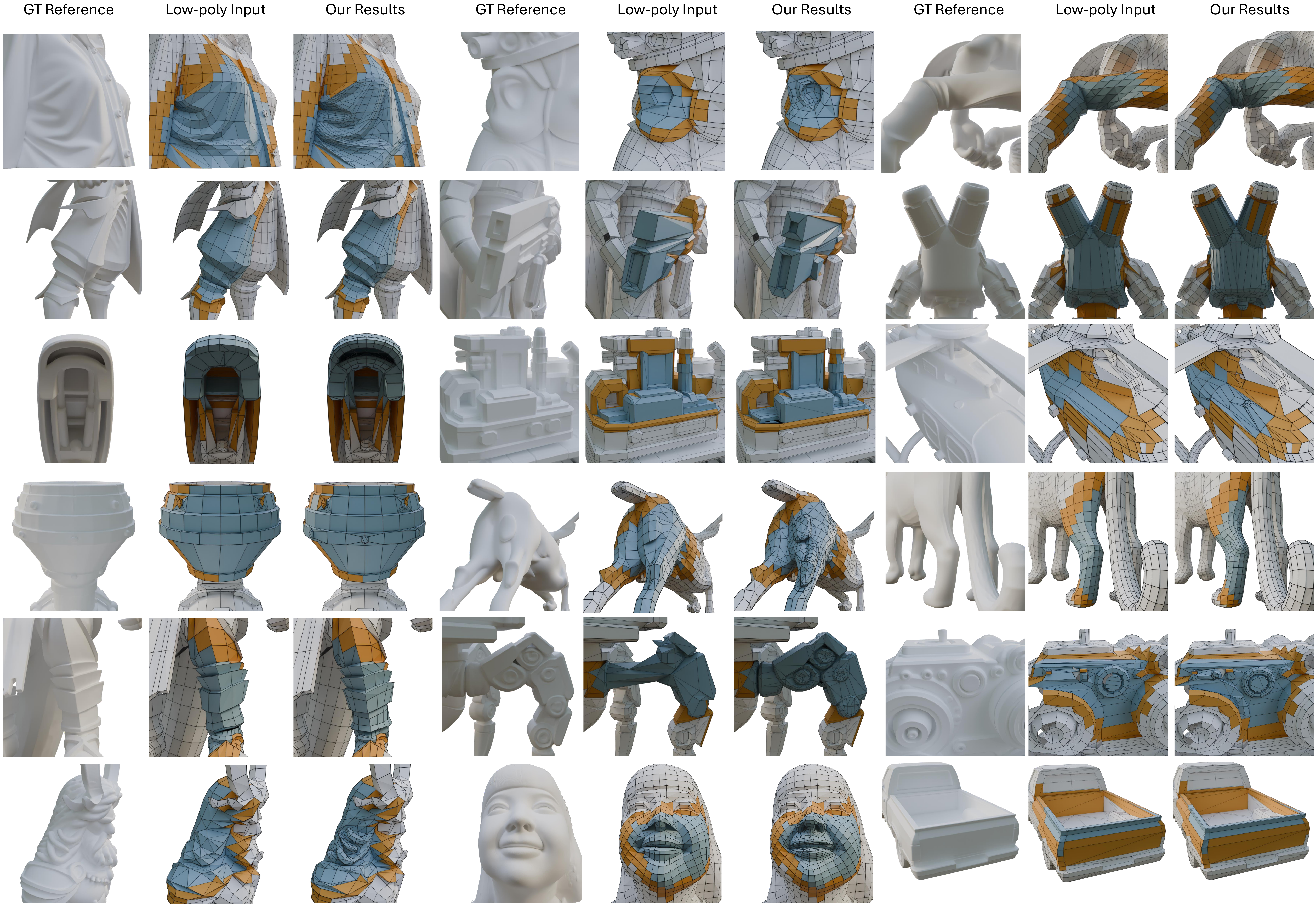}
\caption{More results of our method. 
Our method is able to do accurate local editing on various types of low-poly mesh.}
\label{fig:more-results}
\end{figure}

For more results, please refer to \autoref{fig:more-results}

\end{document}